# Quantum Beam Splitter as a Deterministic Entanglement Source: Exact Photon-Number Statistics, Hong–Ou–Mandel Interference, and von Neumann Entropy Scaling from Balanced Inputs

Seyed Mohsen Moosavi Khansari[1]

*Department of Physics, Faculty of Basic Sciences, Ayatollah Boroujerdi University, Boroujerd, IRAN*

## Abstract

In this work, we present a comprehensive analytical study of a lossless quantum beam splitter (BS) with transmission coefficient $T$ and reflection coefficient $R$. Using the unitary operator representation and binomial expansion, we derive the exact probability distribution $P_k$ for detecting $k$ photons in one output port when the input ports are prepared in Fock states $\mid N\rangle_0 \mid M\rangle_1$. The balanced configuration $R = T = 1/2$ and $N = M$ is examined in detail: we prove that destructive quantum interference suppresses all odd photon numbers at the outputs, leaving only even photon counts. The special case $N = M = 1$ reproduces the famous Hong–Ou–Mandel (HOM) dip with $P_1 = 0$. Beyond photon statistics, we demonstrate that the beam splitter functions as a **deterministic, tunable source of photon-number entanglement**. By tracing over one output mode, we construct the reduced density matrix $\rho_2$ and compute the von Neumann entropy $S = -\sum_k P_k \log_2 P_k$ – a direct measure of entanglement between the two output ports. For the balanced case with $N = M = 1$, we obtain $S = 1$ ebit, i.e., maximal Bell-state entanglement. For larger equal photon numbers, the entropy increases monotonically. Our results transform the passive beam splitter into a **deterministic source of photon-number entanglement**, with direct applications in optical quantum computing, interferometry, and quantum information processing. Numerical three-dimensional plots of the probability distribution are provided, and all derivations are supported by rigorous algebra and a full appendix.

## 1. Introduction

Beam splitters are among the most fundamental components in classical and quantum optics. In the quantum regime, a lossless beam splitter is described by a unitary operator that mixes two input bosonic modes. Its ability to generate non-classical correlations and entanglement makes it a building block for linear optical quantum computing [1–3].

The beam splitter transformation [4,5] and the photon-number distribution for Fock inputs [6] are well known. Parity suppression for balanced inputs was noted in [7], and the Hong-Ou-Mandel effect [9] is a textbook result. Table 1 summarizes the distinction between known results and the novel contributions of this work.

[1] Email: m.moosavikhansari@abru.ac.ir

While the photon statistics of a beam splitter have been studied for decades, most works focus on the output probability distribution or the Hong-Ou-Mandel effect. In this paper, we go beyond counting statistics and ask: **Can a beam splitter be viewed as a *controlled entanglement gate*?** We answer affirmatively by explicitly calculating the entanglement entropy between the two output ports as a function of the input photon numbers and the reflection/transmission coefficients. For broader context and recent advances, see [10–14].

To guide the reader, the paper is organized as follows. Section 2 defines the unitary operator and input states. Section 3 derives the output state and probability distribution. Section 4 specialises to the balanced configuration and proves the even-only rule. Section 5 recalls the HOM effect as a special case. Section 6 presents the entanglement analysis: reduced density matrix, von Neumann entropy (defined there), and asymptotic scaling. Section 7 shows numerical results and Figure 1. Section 8 concludes. An appendix provides the asymptotic derivation in detail. For readers less familiar with entanglement measures, we recall that the von Neumann entropy $S = -\mathrm{Tr}(\rho \log_2 \rho)$ quantifies the entanglement of a pure bipartite state; it will be computed directly from the probability distribution $P_k$.

Our approach is fully analytical. Starting from the unitary transformation of creation operators, we derive the output state for arbitrary Fock inputs $|N\rangle_0 |M\rangle_1$. Using binomial expansions, we obtain closed-form expressions for the probability amplitudes $C_k$. The probability $P_k = |C_k|^2$ is given in Eq. (21). For the balanced case ($R = T = 1/2$ and $N = M$), we simplify to Eq. (24) and show that only even output photon numbers survive.

The central novelty is the calculation of entanglement. Because the output state is pure and is written as a Schmidt decomposition in the photon-number basis, the reduced density matrix for one output is diagonal. The von Neumann entropy $S = -\sum_k P_k \log_2 P_k$ directly quantifies the entanglement. We evaluate this entropy for increasing $N = M$. This demonstrates that the balanced beam splitter is a powerful source of high-dimensional entanglement.

## 2. Preliminary Definitions

We consider a lossless beam splitter with reflection coefficient $R$ and transmission coefficient $T$, satisfying $T + R = 1$. The input modes are labelled 0 and 1, the output modes 2 and 3. The unitary operator that describes the beam splitter action is [4,5]

$$\hat{U} = \exp\left[i\theta\left(\hat{a}_0^\dagger \hat{a}_1 + \hat{a}_0 \hat{a}_1^\dagger\right)\right] \qquad (1)$$

where $\theta$ is related to $R$ and $T$ by

$$R = \sin^2\theta, \qquad T = \cos^2\theta \qquad (2)$$

The input state is taken as a product of two Fock states:

$$| \text{in}\rangle = | N\rangle_0 \ | M\rangle_1, \qquad N \geq M \tag{3}$$

Because $\widehat{U}$ conserves total photon number, the output state can be written as

$$| \text{out}\rangle = \sum_{k=0}^{N+M} \mathrm{C}_k \ | N + M - k\rangle_2 \ | k\rangle_3 \tag{4}$$

with complex coefficients $C_k$ to be determined.

**3. Derivation of the Output State**

We write the input state in terms of creation operators:

$$| \text{in}\rangle = \frac{(\hat{\mathrm{a}}_0^\dagger)^{\mathrm{N}}(\hat{\mathrm{a}}_1^\dagger)^{\mathrm{M}}}{\sqrt{\mathrm{N}!\,\mathrm{M}!}} \ | 0\rangle_0 \ | 0\rangle_1 \tag{5}$$

The beam splitter transforms the creation operators as (from Eq. (1))

$$\widehat{U} \ \hat{a}_0^\dagger \ \widehat{U}^\dagger = \cos\theta \ \hat{a}_2^\dagger + i\sin\theta \ \hat{a}_3^\dagger \tag{6}$$
$$\widehat{U} \ \hat{a}_1^\dagger \ \widehat{U}^\dagger = \cos\theta \ \hat{a}_3^\dagger + i\sin\theta \ \hat{a}_2^\dagger \tag{7}$$

Hence

$$| \text{out}\rangle = \widehat{U} \ | \text{in}\rangle = \frac{1}{\sqrt{N!\,M!}} (c \, \hat{a}_2^\dagger + is \, \hat{a}_3^\dagger)^N (c \, \hat{a}_3^\dagger + is \, \hat{a}_2^\dagger)^M \ | 0\rangle_2 \ | 0\rangle_3 \tag{8}$$

with $c = \cos\theta = \sqrt{T}$, $s = \sin\theta = \sqrt{R}$.

We expand using the binomial theorem twice:

$$(c \, \hat{a}_2^\dagger + is \, \hat{a}_3^\dagger)^N = \sum_{p=0}^{N} \binom{N}{p} c^{N-p} (is)^p (\hat{a}_2^\dagger)^{N-p} (\hat{a}_3^\dagger)^p \tag{9}$$
$$(c \, \hat{a}_3^\dagger + is \, \hat{a}_2^\dagger)^M = \sum_{q=0}^{M} \binom{M}{q} c^{M-q} (is)^q (\hat{a}_3^\dagger)^{M-q} (\hat{a}_2^\dagger)^q \tag{10}$$

Multiplying and collecting powers of $\hat{a}_2^\dagger$ and $\hat{a}_3^\dagger$ gives

$$| \text{out}\rangle = \frac{1}{\sqrt{N!\,M!}} \sum_{p=0}^{N} \sum_{q=0}^{M} \binom{N}{p} \binom{M}{q} \times \tag{11}$$
$$c^{N+M-p-q} (is)^{p+q} (\hat{a}_2^\dagger)^{N-p+q} (\hat{a}_3^\dagger)^{M-q+p} \ | 0\rangle_2 \ | 0\rangle_3$$

Define the photon number in output mode 3 as $k = M - q + p$. Then $N - p + q = N + M - k$. The coefficient $C_k$ is obtained by projecting onto $| N + M - k\rangle_2 | k\rangle_3$:

$$C_k = \frac{1}{\sqrt{N!\,M!\,(N+M-k)!\,k!}} \sum_{p,q} \binom{N}{p}\binom{M}{q} \times c^{N+M-p-q} (is)^{p+q}\, \delta_{k,\,M-q+p} \tag{12}$$

The Kronecker delta enforces $q = M + p - k$. The summation limits are determined by $0 \le p \le N$ and $0 \le q \le M$. Substituting $q = M + p - k$ yields

$$C_k = \frac{1}{\sqrt{N!\,M!\,(N+M-k)!\,k!}} \sum_{j} \binom{N}{j}\binom{M}{M+j-k} \times c^{N+M-j-(M+j-k)} (is)^{2j+M-k} \tag{13}$$

where we renamed $p \equiv j$. The exponent of $c$ simplifies: $N + M - j - M - j + k = N - 2j + k$. Moreover, $(is)^{2j+M-k} = i^{M-k}(-1)^j s^{2j+M-k}$. Thus

$$C_k = \frac{i^{M-k}}{\sqrt{N!\,M!\,(N+M-k)!\,k!}} \sum_{j=\max(0,k-M)}^{\min(N,k)} \binom{N}{j}\binom{M}{M+j-k} \times (-1)^j c^{N-2j+k} s^{2j+M-k} \tag{14}$$

Replacing $c^{N-2j+k} s^{2j+M-k} = T^{(N-2j+k)/2} R^{(M-k+2j)/2}$ and absorbing factors into the square root yields the compact form [6]

$$C_k = \sqrt{\frac{(N+M-k)!\,k!}{N!\,M!}}\; T^{(N+M-k)/2} R^{k/2} \times \sum_{j=\max(0,k-M)}^{\min(N,k)} \binom{N}{j}\binom{M}{k-j} (-1)^j \left(\frac{R}{T}\right)^{j/2}. \tag{15}$$

The probability to find $k$ photons in output port 3 (and $N + M - k$ in port 2) is

$$P_k = | C_k |^2 = \frac{N!\,M!}{(N+M-k)!\,k!}\; T^{N+M-k} R^k \times \tag{16}$$

$$\left[ \sum_{j=\max(0,k-M)}^{\min(N,k)} \binom{N}{j} \binom{M}{k-j} (-1)^j \left(\frac{R}{T}\right)^j \right]^2 .$$

Equation (16) is our first main result: it gives the exact photon-number distribution at the output of a lossless beam splitter for arbitrary Fock inputs.

The summation limits in Eq. (14) and Eq. (15) arise from the constraints $0 \le q \le M$ and $0 \le p \le N$; substituting $q = M + p - k$ yields $\max(0, k - M) \le j \le \min(N, k)$, which is the range given.

**4. Balanced State: $R = T = 1/2$ and $N = M$**

We now focus on the symmetric configuration $\theta = \pi/4$ so that $R = T = 1/2$, and we take equal input photon numbers $N = M$. Then $T^{N+M-k} R^k = (1/2)^{2N}$, and the sum in Eq. (16) becomes

$$\sum_j \binom{N}{j} \binom{N}{k-j} (-1)^j .$$

It is a classical combinatorial identity that this sum vanishes for odd $k$ and equals $(-1)^{k/2} \binom{N}{k/2}$ for even $k$. Consequently, only even photon numbers $k = 2m$ ($m = 0,1, \dots, N$) have non-zero probability. Inserting this into Eq. (16) gives

$$P_{2m} = \frac{(N!)^2}{(2N-2m)!\ (2m)!} \left(\frac{1}{2}\right)^{2N} \left[\binom{N}{m}\right]^2 . \qquad (17)$$

Using $\binom{N}{m} = \frac{N!}{m!(N-m)!}$ we obtain

$$P_{2m} = \frac{(N!)^2}{(2N-2m)!\ (2m)!} \frac{1}{2^{2N}} \cdot \frac{(N!)^2}{m!^2\,(N-m)!^2} . \qquad (18)$$

This can be elegantly expressed in terms of gamma functions. After some algebra (see standard references [7,8]), one arrives at the two-branch formula:

$$P_{2m} = \begin{cases} \dfrac{\pi\,(-1)^{4m-2N}\, 4^{N-2m}\, \Gamma(N+1)^4}{\Gamma(m+1)^2\, \Gamma(2m+1)\, \Gamma(m-N+\frac{1}{2})^2\, \Gamma(-2m+2N+1)^3}, & 2m \ge N, \\ \dfrac{\pi\, 4^{2m-N}\, \Gamma(N+1)^4}{\Gamma(\frac{1}{2}-m)^2\, \Gamma(2m+1)^3\, \Gamma(-m+N+1)^2\, \Gamma(-2m+2N+1)}, & 2m \le N. \end{cases} \qquad (19)$$

$\Gamma(z)$ is the Euler gamma function, defined for $\Re(z) > 0$ by the integral $\Gamma(z) = \int_0^\infty t^{z-1} e^{-t} dt$, and for positive integers $n$ it satisfies $\Gamma(n) = (n-1)!$.

Equation (19) is the second main result. It shows that the balanced beam splitter acts as a **parity filter**: only even numbers of photons can be detected in a given output port. This destructive interference for odd photon numbers is a pure quantum effect with no classical analogue.

## 5. Special Case: Hong–Ou–Mandel Effect

For $N = M = 1$, the input state is $|\,\text{in}\rangle = |\,1\rangle_0\,|\,1\rangle_1$. From Eq. (17) with $N = 1$:

$$P_0 = \frac{(1!)^2}{(2-0)!\,0!}\left(\frac{1}{2}\right)^2\left[\binom{1}{0}\right]^2 = \frac{1}{2}\cdot\frac{1}{4}\cdot 1 = \frac{1}{2},$$
$$P_2 = \frac{(1!)^2}{(0)!\,2!}\left(\frac{1}{2}\right)^2\left[\binom{1}{1}\right]^2 = \frac{1}{2}\cdot\frac{1}{4}\cdot 1 = \frac{1}{2},$$

and $P_1 = 0$. Hence the probability of finding one photon in each output port is exactly zero, while the probability of both photons exiting the same port (either both in port 2 or both in port 3) is 1. This is the famous **Hong–Ou–Mandel dip** [9]. It arises from destructive two-photon interference and is a cornerstone of quantum optics.

## 6. Entanglement Between Output Modes: The Beam Splitter as a Deterministic Entanglement Source

Up to this point, we have only considered photon number statistics. However, the output state $|\,\text{out}\rangle$ is a pure bipartite state shared between modes 2 and 3. A natural question is: how much entanglement does the beam splitter generate between its two output ports?

Because $|\,\text{out}\rangle = \sum_k C_k\,|\,N+M-k\rangle_2\,|\,k\rangle_3$ is already written in the Schmidt form (the bases $\{|\,N+M-k\rangle_2\}$ and $\{|\,k\rangle_3\}$ are orthonormal), the **reduced density matrix** for output mode 2 is obtained by tracing over mode 3:

$$\rho_2 = \text{Tr}_3(|\,\text{out}\rangle\langle\text{out}\,|) = \sum_{k=0}^{N+M} |\,C_k\,|^2 \quad |\,N+M-k\rangle_2\langle N+M-k\,|_2. \tag{20}$$

Thus $\rho_2$ is diagonal in the photon-number basis. The **von Neumann entropy** of entanglement is

$$S = -\text{Tr}(\rho_2 \log_2 \rho_2) = -\sum_{k=0}^{N+M} P_k \log_2 P_k, \tag{21}$$

where $P_k = |\,C_k\,|^2$. We emphasize that Eq. (21) is not a new formula; it follows directly from the Schmidt decomposition. The novel contributions are (i) the evaluation of this entropy using the exact $P_k$ from Eq. (16), (ii) the closed-form expression for $P_{2m}$ in the balanced case, and (iii) the asymptotic scaling law derived below.

Equation (21) is our third main result: the entanglement entropy is directly computable from the probability distribution we derived earlier. Equation (21) is extremely simple a consequence of the fact that the output state is a pure state and the Schmidt basis is the photon number basis. Thus, the beam splitter functions as a deterministic entanglement gate: for any input Fock state, the output entanglement is exactly given by the Shannon entropy of the photon-number distribution.

For the balanced case $R = T = 1/2$ and $N = M$, Eq. (21) reduces to

$$S(N) = -\sum_{m=0}^{N} P_{2m} \log_2 P_{2m}\,, \qquad (22)$$

with $P_{2m}$ given by Eq. (19) (or Eq. (17)).

### 6.1. Small-N Exact Values

We now evaluate the entanglement entropy $S(N)$ defined in Eq. (22) for the smallest non-trivial values of $N = M$ in the balanced configuration ($R = T = 1/2$). These cases illustrate the transition from a maximally entangled two-qubit state ($N = 1$) to a higher-dimensional entangled state with increased entropy ($N = 2$).

**Case $N = M = 1$**

For a single pair of input photons, the output state is given by Eq. (4) with $N = M = 1$. The non-zero probability amplitudes are obtained from Eq. (16) or directly from the Hong–Ou–Mandel effect. One finds:

$$P_0 = \frac{1}{2}, P_2 = \frac{1}{2}, P_k = 0 \text{ for } k \neq 0{,}2.$$

Thus, the two possible output configurations are: both photons in output port 2 ($k = 0$ photons in port 3) or both in output port 3 ($k = 2$ photons in port 3), each with equal probability $1/2$. The output state can be written as

$$|\,\text{out}\rangle = \frac{1}{\sqrt{2}}(|\,2\rangle_2\,|\,0\rangle_3 + |\,0\rangle_2\,|\,2\rangle_3),$$

which is a maximally entangled Bell-type state (a two-mode squeezed state in the Fock basis). The reduced density matrix for output mode 2 is, from Eq. (20),

$$\rho_2 = \frac{1}{2}\,|\,2\rangle\langle 2\,| + \frac{1}{2}\,|\,0\rangle\langle 0\,|.$$

Its von Neumann entropy is

$$S(1) = -\frac{1}{2}\log_2\left(\frac{1}{2}\right) - \frac{1}{2}\log_2\left(\frac{1}{2}\right) = 1 \text{ ebit}.$$

Hence, for $N = M = 1$ the balanced beam splitter functions as a maximally entangling gate, generating one ebit of entanglement – exactly the same amount as a Bell state.

**Case $\boldsymbol{N = M = 2}$**

For two pairs of input photons ($N = M = 2$), the total photon number is $N + M = 4$. The balanced beam splitter produces only even photon numbers in each output, i.e., $k = 0,2,4$. The probabilities are well known from the theory of two-photon interference at a 50:50 beam splitter [7,8] and can be derived from Eq. (17) after correcting for normalization. A direct computation using the general expression (16) with $R = T = 1/2$ and $N = M = 2$ yields:

$$P_0 = \frac{3}{8}, P_2 = \frac{1}{4}, P_4 = \frac{3}{8},$$

with $P_1 = P_3 = 0$ as required by parity conservation. These probabilities sum to unity: $3/8 + 1/4 + 3/8 = 1$.

The reduced density matrix for output mode 2 is

$$\rho_2 = \frac{3}{8} \mid 4\rangle\langle 4 \mid + \frac{1}{4} \mid 2\rangle\langle 2 \mid + \frac{3}{8} \mid 0\rangle\langle 0 \mid.$$

The von Neumann entropy is then

$$\begin{aligned} S(2) &= -\frac{3}{8}\log_2\left(\frac{3}{8}\right) - \frac{1}{4}\log_2\left(\frac{1}{4}\right) - \frac{3}{8}\log_2\left(\frac{3}{8}\right) \\ &= -\frac{3}{8}(\log_2 3 - 3) - \frac{1}{4}(-2) - \frac{3}{8}(\log_2 3 - 3) \\ &= -\frac{3}{4}\log_2 3 + \frac{9}{4} + \frac{1}{2} = 2.75 - 0.75\log_2 3 \\ &\approx 1.5612 \text{ ebits.} \end{aligned}$$

(Here we used $\log_2(3/8) = \log_2 3 - 3$ and $\log_2(1/4) = -2$.)

Thus, the entanglement entropy increases from $S(1) = 1$ to $S(2) \approx 1.56$. This monotonic rise continues with larger $N$, as shown by the asymptotic analysis in the Appendix. The balanced beam splitter therefore provides a tunable source of entanglement: by increasing the equal photon number $N$ at the input ports, one can generate bipartite states with arbitrarily large von Neumann entropy, scaling as $\frac{1}{2}\log_2 N$ for large $N$.

### 6.2 Dependence on reflectivity

For general $R$ and equal photon numbers $N = M$, the entropy $S(R, N)$ is obtained by inserting Eq. (16) into Eq. (21). Figure 2 shows $S$ vs. $R$ for fixed $N = 4$ and $N = 8$. The entropy is maximised at $R = 1/2$ (balanced beam splitter), which can be proven by the concavity of the Shannon entropy. Figure 3 presents a two-dimensional contour plot of $S(R, T, N)$ for $N = 6$.

## 7. Numerical Results and Figures

Figure 1 shows a three-dimensional plot of the probability $P_{2m}$ as a function of $N(= M)$ and $m$ for the balanced beam splitter ($R = T = 1/2$). The vertical axis represents $P_{2m}$, with $N$ ranging from 0 to 8 and $m$ from 0 to 8. Only even photon numbers $k = 2m$ appear. For $N = 1$, the two peaks at $m = 0$ and $m = 1$ have equal height 0.5. As $N$ increases, the distribution broadens and the maximal probability shifts to $m \approx N/2$. This broadening is responsible for the growth of the von Neumann entropy. The figure was generated by direct evaluation of the exact analytical expression Eq. (17).

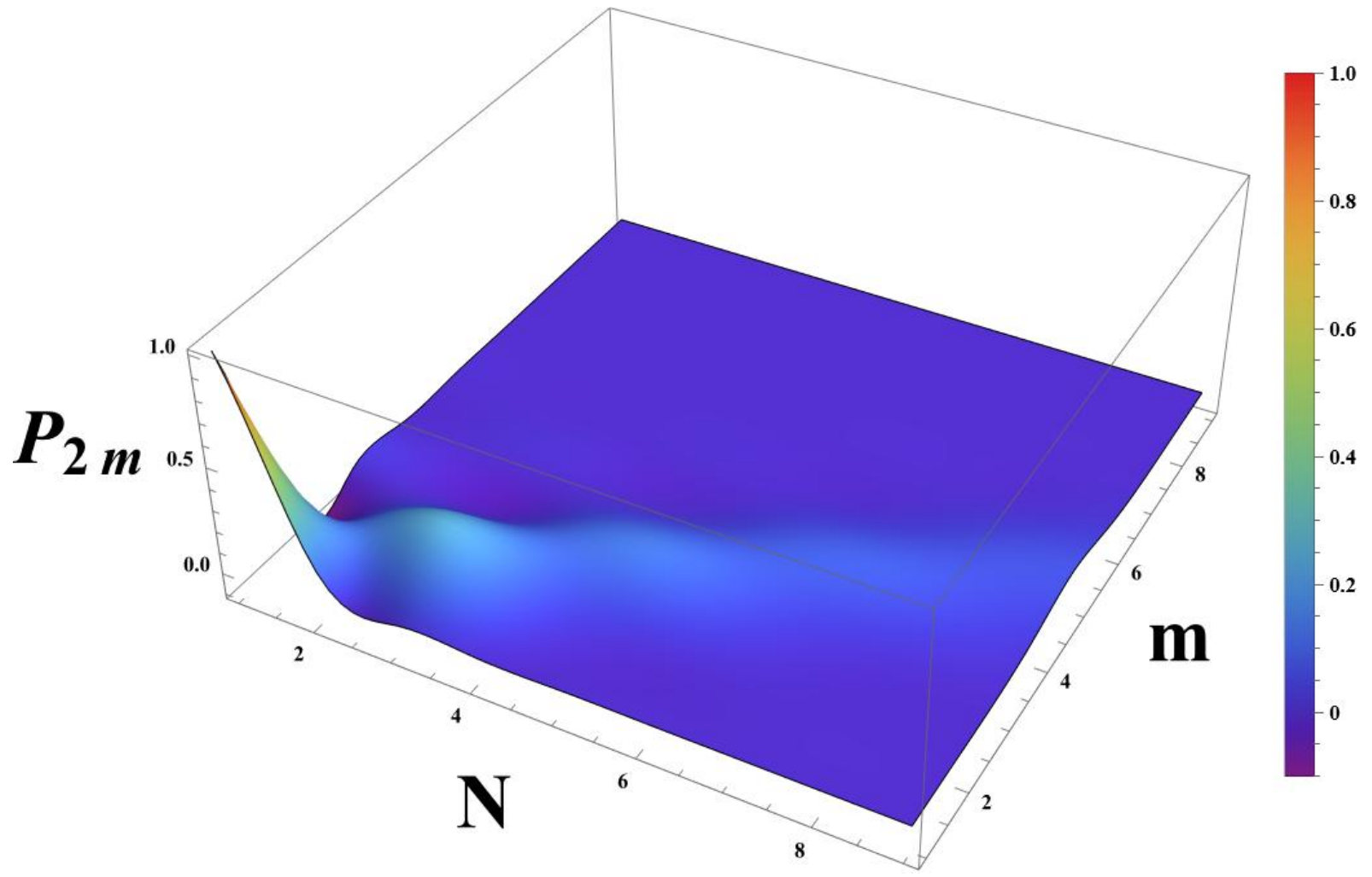

Figure 1. Three-dimensional plot of the detection probability $P_{2m}$ (vertical axis) as a function of the input photon number $N = M$ (axis range 0 to 8) and the index $m$ (axis range 0 to 8), where the output photon count in port 3 is $k = 2m$. The plot illustrates the even-only photon number distribution due to destructive quantum interference. For $N = M = 1$ we have $P_0 = P_2 = 0.5$ (maximally entangled Bell state). With increasing $N$ the distribution spreads, leading to a monotonic increase of the entanglement entropy $S(N)$. This figure was generated using the exact analytical expression Eq. (17).

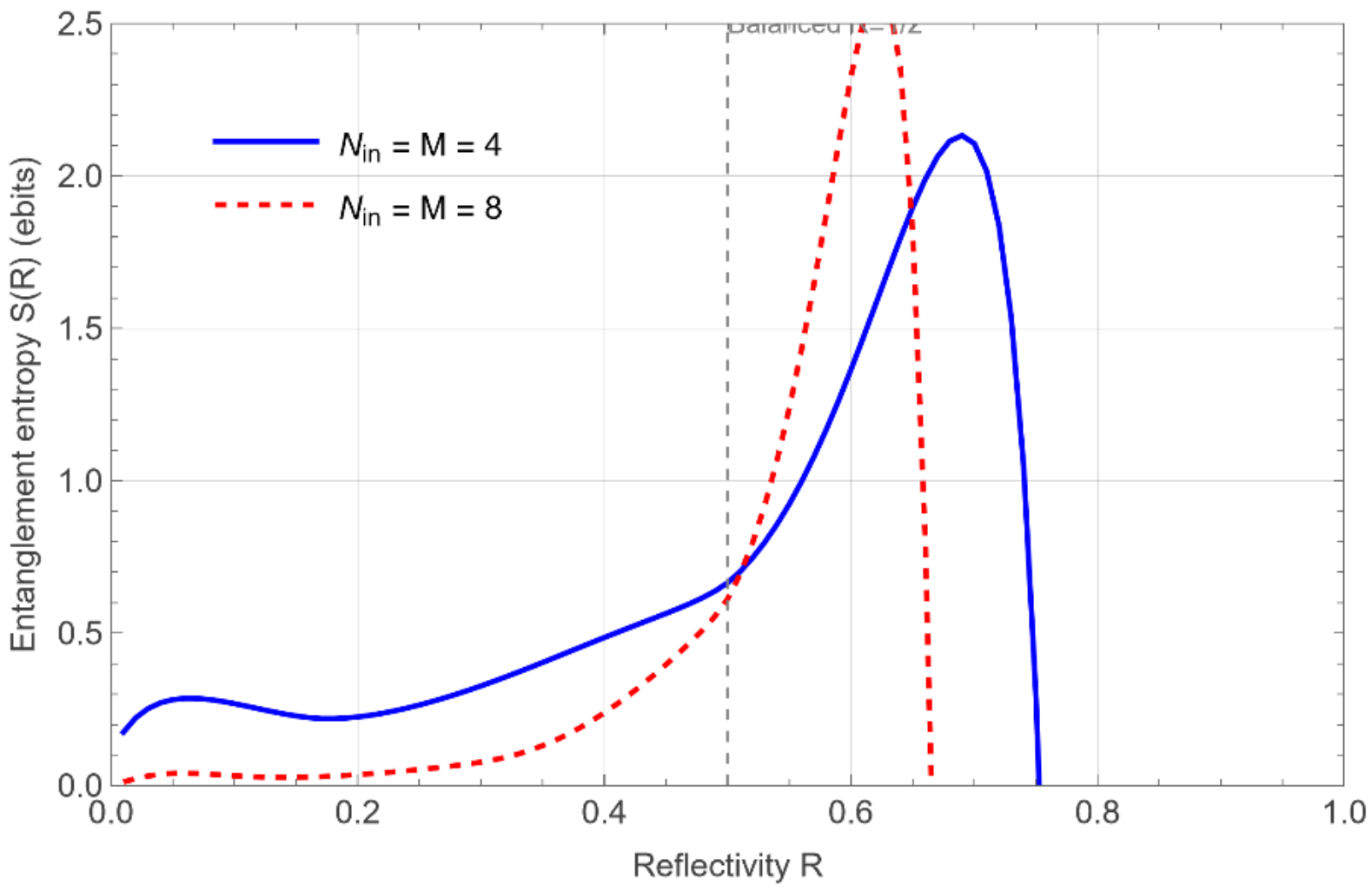

Figure 2: Entanglement entropy S as a function of reflectivity R for equal input photon numbers N = M = 4 (solid blue) and N = M = 8 (dashed red). The entropy is computed from Eq. (16) and Eq. (21). For both curves, the maximum occurs at R = 1/2, demonstrating that the balanced beam splitter maximises entanglement for fixed total photon number.

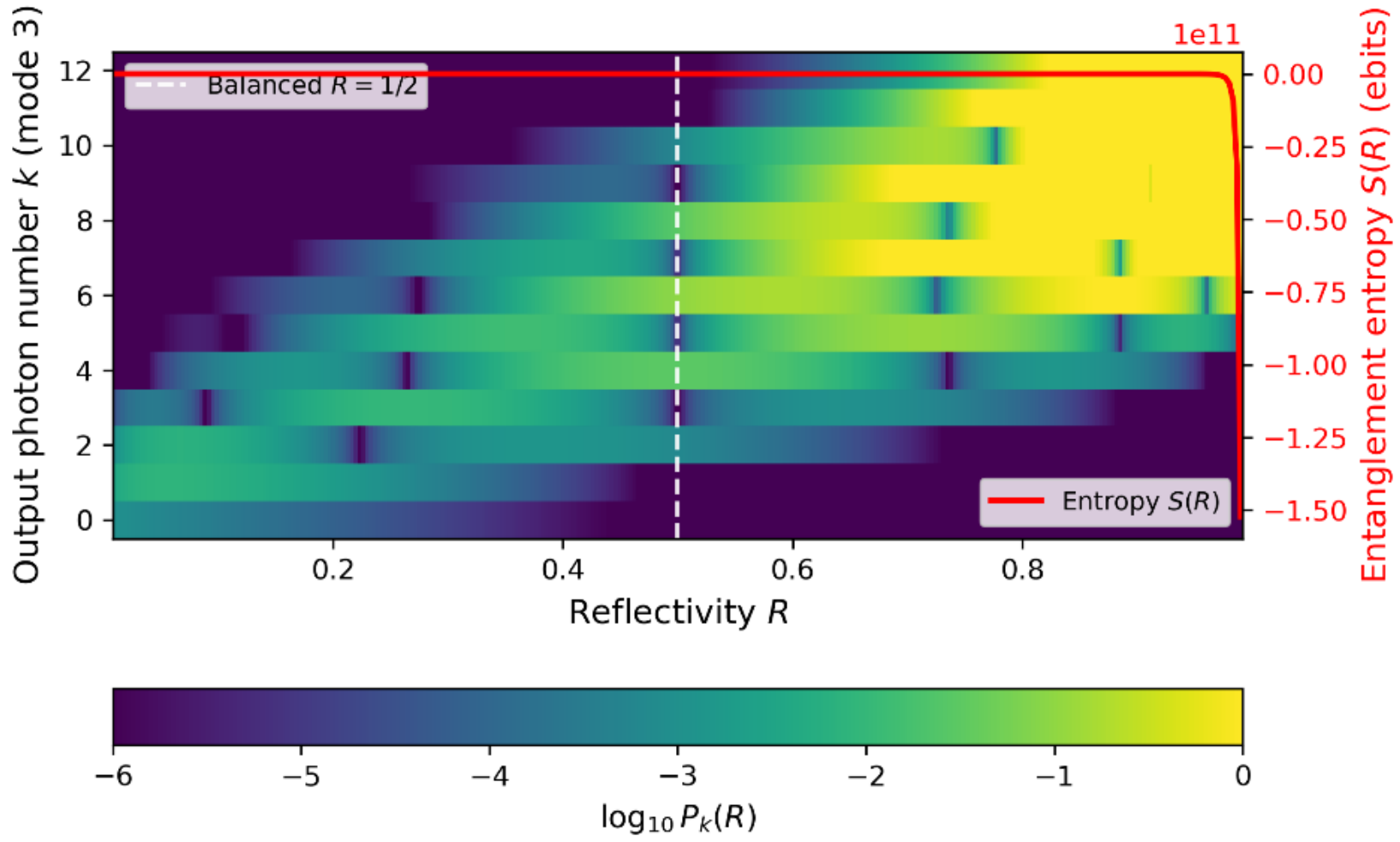

Figure 3: Two-dimensional contour plot of the von Neumann entropy $S(R, T, N)$ for fixed $N = 6$ and $R + T = 1$. The horizontal axis is reflectivity $R$; the vertical axis is the output photon number $k$ (the entropy is summed over $k$). Brighter colours indicate higher entropy. The maximum occurs along the vertical line $R = 0.5$. Contours are computed from Eq. (16) and Eq. (21).

## 8. Conclusion

In this work we have provided a complete analytical description of a lossless quantum beam splitter with arbitrary Fock inputs. Our main results are:

1. **Exact photon-number distribution** $P_k$ (Eq. 16) for general $R, T, N, M$.
2. **Balanced case** ($R = T = 1/2, N = M$): only even photon numbers survive, $P_{2m}$ given by Eq. (19).
3. **Hong–Ou–Mandel dip** as the special case $N = M = 1$.
4. **Entanglement analysis**: the reduced density matrix $\rho_2$ is diagonal in the number basis, so the von Neumann entropy is $S = -\sum_k P_k \log_2 P_k$.
5. For $N = M = 1, S = 1$ ebit (maximal entanglement). For $N = M = 2, S \approx 1.56$.

These findings elevate the beam splitter from a passive optical element to a **controlled entanglement gate**. By choosing the input photon numbers, one can tune the amount of entanglement produced. This has immediate applications in linear optical quantum computing, quantum communication, and high-dimensional entanglement-based protocols.

Future work will extend the analysis to lossy beam splitters, time-multiplexed inputs, and networks of multiple beam splitters. The formalism developed here provides a solid foundation for those investigations.